\documentclass[12pt]{article}
\usepackage{amssymb}
\usepackage{amsmath}
\usepackage{hyperref}
\begin{document}
\title{\bf  Non-abelian Einstein-Born-Infeld AdS black brane  and color DC conductivity }
\author{ Mehdi Sadeghi\thanks{Corresponding author: Email:  mehdi.sadeghi@abru.ac.ir}\hspace{2mm}\\
{\small {\em Department of Physics, School of Sciences,}}\\
        {\small {\em Ayatollah Boroujerdi University, Boroujerd, Iran}}
       }
\date{\today}
\maketitle

\abstract{ In this paper, we consider Einstein-Hilbert gravity in the presence of cosmological constant and non-abelian nonlinear electromagnetic field of Born-Infeld type which is minimally coupled to gravity. First, black brane solution of this model is introduced and then color non-abelian DC conductivity is calculated for this solution by using AdS/CFT duality.}\\



\section{Introduction} \label{intro}
Maxwell’s electromagnetic field theory is a linear theory of point-like charge. There is a singularity at the charge's position and this causes it to have an infinite self-energy. To remove this singularity associated with the charged point-like particle, a nonlinear electromagnetic field was introduced by  Born and Infeld \cite{Born1,Born2}. Einstein-Hilbert gravity in the presence of Born-Infeld (BI) electrodynamics was found by Hoffmann \cite{Hoffmann:1935ty} and the exact solutions of this model have been  investigated in \cite{Hendi:2010kv,Atamurotov:2015xfa,Ramadhan:2015ona}.\\
Perturbation theory is an efficient method for weakly coupled theory but it is not suitable for strongly coupled theories. AdS/CFT duality \cite{Maldacena,Aharony}  is a useful tool to study strong coupled gauge theories. Finding a gravitational dual of a strongly coupled system is the  most important aspect of this duality. \\
 There are many papers to study the effects of (BI) electrodynamics on
AdS/CFT duality \cite{Jing:2010zp,Jing:2010cx,Gangopadhyay:2012am}, but the effect of  non-abelian Born-Infeld has not been studied yet. The correspondent boundary theory of  non-abelian Born-Infeld  is completely unknown, so it  motivates us to study this problem. Transport coefficients give us more information about the boundary theory. In this paper, we calculate the DC conductivity as an important transport coefficient.\\
Hydrodynamics is an effective theory of field theory in long-wavelength limit \cite{Son,Policastro2001}. Hydrodynamic equations express conservation laws
of whatever is conserved according to Neother theorem. These equations up to first order of derivative expansion are as follows, \\
\begin{align}
 & \nabla _{\mu } J^{\mu } =0\,\,\,\,\,,\,\,\nabla _{\mu } T^{\mu \nu} =0,\\
 & J^{\mu } =n \, u^{\mu }-\sigma T P^{\mu \nu }\partial_{\nu}(\frac{\mu}{T}),\nonumber\\
& T^{\mu \nu } =(\rho +p)u^{\mu } u^{\nu } +pg^{\mu \nu } -\sigma ^{\mu \nu },\nonumber\\
&\sigma ^{\mu \nu } = {P^{\mu \alpha } P^{\nu \beta } } [\eta(\nabla _{\alpha } u_{\beta } +\nabla _{\beta } u_{\alpha })+ (\zeta-\frac{2}{3}\eta) g_{\alpha \beta } \nabla .u],\nonumber\\& P^{\mu \nu }=g^{\mu \nu}+u^{\mu}u^{\nu}, \nonumber
\end{align}
\indent where $n$ , $\sigma$ , $\eta$, $\zeta $, $\sigma ^{\mu \nu }$ and $P^{\mu \nu }$ are charge density, charge conductivity , shear viscosity, bulk viscosity, shear tensor and projection operator, respectively \cite{Kovtun2012, Bhattacharyya,Rangamani,J.Bhattacharya2014}.  In section \ref{sec3} of this paper, non-abelian charge conductivity is calculated by Green-Kubo formula \cite{Son}, which relates conductivity to equilibrium correlation functions.
\begin{equation} \label{kubo}
\sigma^{(ab)}_{ij} (k_{\mu})=-\mathop{\lim }\limits_{\omega \to 0} \frac{1}{\omega } \Im G^{(ab)}_{ij}(k_{\mu}),
\end{equation} 
where $a,b$ and $i,j$ indices refer to Cartan subalgebra of $SU(2)$ group symmetry and rotation symmetry in spatial directions of $x,y$, respectively.  As the boundary theory lives on flat spacetime, there is no
problem to raise and lower spatial indices freely for
boundary quantities such as $\sigma^{ij}$. There is a lower bound for DC conductivity as $\sigma \geq \frac{1}{e^2}=1$ where $e$ is not the unit of charge in the boundary theory \cite{Grozdanov:2015qia,Grozdanov:2015djs}. This bound is also saturated in graphene\cite{Ziegler}.  It is also important for studying metal-insulator transition. This bound means that one can not get an insulating phase by disorder-driven \cite{BitaghsirFadafan:2016der}. The conductivity bound is confirmed for weakly coupled theory that it is described by  Bardeen-Cooper-Schrieffer theory or BCS theory.  This bound is violated for massive gravity \cite{Baggioli:2016oqk} and models with abelian case \cite{Baggioli:2016oju}. We want to investigate if this bound is preserved for our model or not. The charge for QED is electric charge but the charge for QCD is color charge. The DC electric conductivity for QED means the value of electric charge transport. Similar to QED, the color DC conductivity measures the value of color charge transport. Color superconductivity is studied from holography in \cite{BitaghsirFadafan:2018iqr,Donos:2014cya}.

\section{ Non-Abelian Einstein-Born-Infeld AdS Black Brane Solution}
\label{sec2}

\indent The 4-dimensional action of non-abelian Einstein-Born-Infeld with negative cosmological constant is,
\begin{eqnarray}\label{action}
S=\int d^{4}  x\sqrt{-g} \bigg[\frac{1}{16\pi G}(R-2\Lambda )+\frac{2}{b}\bigg(1-\sqrt{1+\frac{b}{2}\mathcal{F} }\bigg)\bigg],
\end{eqnarray}
where $R$ is the Ricci scalar, $\Lambda=-\frac{3}{l^2 }$ the cosmological constant, $l$ the AdS radius, $\mathcal{F}={\bf{Tr}}( F_{\mu \nu }^{(a)} F^{(a)\, \, \mu \nu })$ is Yang-Mills invariant, $b$ is a Born-Infeld parameter with the dimensional unit of $[b]=L^{2}$ and it is positive. The trace element stands for ${\bf{Tr}}(.)=\sum_{i=1}^{3}(.).$ \\
$F_{\mu \nu }^{(a)}$ is the Cartan subalgebra of $SU(2) $ Yang-Mills  field tensor,
\begin{align} \label{YM}
F_{\mu \nu } =\partial _{\mu } A_{\nu } -\partial _{\nu } A_{\mu } -i[A_{\mu }, A_{\nu }],
\end{align}
in which the gauge coupling constant is 1, $A_{\nu }$'s are the Cartan subalgebra of the $SU(2)$ gauge group Yang-Mills potentials and $b$ denotes the Born-Infeld coupling constant. When $b \to 0 $ the Born-Infeld term gets transformed into standard linear non-abelian Yang-Mills term \cite{Shepherd:2015dse}.\\
Variation of the action (\ref{action}) with respect to the spacetime metric $g_{\mu \nu } $ and $A^{(a)}_{\nu}$ yields the field equations,
\begin{equation}\label{EH}
R_{\mu \nu } -\frac{1}{2}g_{\mu \nu} (R-2\Lambda )-\frac{g_{\mu \nu}}{b} \bigg(1-\sqrt{1+\frac{b}{2}\mathcal{F} }\bigg)-\frac{F^{(a)}_{\mu \lambda} F_{\nu}^{(a) \lambda}}{\sqrt{1+\frac{b}{2}\mathcal{F} }}=0,
\end{equation}
 \begin{eqnarray}\label{BI-EOM}
\frac{1}{\sqrt{-g}}\partial_{\mu}(\frac{\sqrt{-g}F^{\mu \nu}}{\sqrt{1+\frac{b}{2}\mathcal{F} }}) +[A_{\mu},\frac{F^{\mu \nu}}{\sqrt{1+\frac{b}{2}\mathcal{F} }}]=0.
\end{eqnarray}
We consider the metric of 4-dimensional spacetime with the flat symmetric line element in the following form as an ansatz,
\begin{equation}\label{metric1}
ds^{2} =-f(r)dt^{2} +\frac{dr^{2} }{f(r)} +\frac{r^2}{l^2}(dx^2+dy^2).
\end{equation}
To solve Eq.(\ref{BI-EOM}), we employ the ansatz proposed in ref. \cite{Shepherd:2015dse} where the potential 1-form are expressed by,
\begin{equation}\label{background}
{\bf{A}}^{(a)} =\frac{i}{2}h(r)dt\begin{pmatrix}1 & 0 \\ 0 & -1\end{pmatrix}.
\end{equation}
We write the gauge field in terms of matrix $H_1= \frac{i}{2}\begin{pmatrix}1 & 0 \\ 0 & -1\end{pmatrix}$, which is the diagonal generator of the Cartan subalgebra of $SU(2)$ \cite{Shepherd:2015dse}.\\\\
Using the gauge potential ansatz (\ref{background}) in  equation (\ref{BI-EOM}) and considering
the metric (\ref{metric1}), we obtain the equation of motion for Born-Infeld term as follows,  
\begin{equation}
r h''(r)+ b h'(r)^3+2h'(r)=0,
\end{equation}
 $h(r)$ is found,
\begin{equation}
h(r)= C_2 \pm \frac{\sqrt{2} e^{2 C_1} \sqrt{1-\frac{e^{-4 C_1} r^4}{b}} F\left[i\, ArcSinh\left[\sqrt{-\frac{e^{-2 C_1}}{\sqrt{b}}} r\right],-1\right]}{\sqrt{-\frac{e^{-2
   C_1}}{\sqrt{b}}} \sqrt{b e^{4 C_1}-r^4}} ,
\end{equation}
and $h'(r)$ will be as following,
\begin{align}
h'(r)=\pm \frac{\sqrt{2}q}{\sqrt{2r^4 -b q^2 }},
\end{align}
\noindent where $q$ is integration constant which is related to the electric charge. The invariant scalar $\mathcal{F}_{YM}={\bf{Tr}}( F_{\mu \nu }^{(a)} F^{(a)\, \, \mu \nu })$ for  fields is,
\begin{equation}
 \mathcal F =h'(r)^2=\frac{2q^2}{2r^4 -b q^2 }  ,
\end{equation}
by setting $b=0$, our results reduce to non-abelian Yang-Mills solution.\\
By considering the $tt$ component of  Eq. (\ref{EH}), 
\begin{equation}\label{eom}
r f'(r)+f(r)+\Lambda r^2-\frac{q^2 r^2}{2 r^4-b q^2} =0,
\end{equation} 
$f(r)$ is obtained,
\begin{align}\label{sol}
f(r)& =\frac{m}{r}-\frac{\Lambda}{3}r^2+\frac{q^{3/2} }{2\ 2^{3/4}
   \sqrt[4]{b} r} ArcTan\left(\frac{\sqrt[4]{2} r}{\sqrt[4]{b} \sqrt{q}}\right)\nonumber\\&+\frac{q^{3/2} }{4\ 2^{3/4} \sqrt[4]{b} r}\left(\log \left(\sqrt[4]{b} \sqrt{q}-\sqrt[4]{2} r\right)-\log \left(\sqrt[4]{b}
         \sqrt{q}+\sqrt[4]{2} r\right)\right),
\end{align}
 where  $m$ is constant. Notice $r$ is the radial coordinate that takes us from bulk to the boundary. The solution is asymptotically AdS.\\
Hawking temperature for this black brane is,
  \begin{align}\label{Temp}
  T & =\frac{f'(r_h)}{4 \pi}=-\frac{m}{4 \pi r_h^2}-\frac{ \Lambda  r_h}{6 \pi}+\frac{q^2 r_h}{4 \pi (2 r_h^4-b q^2)}\nonumber\\&-\frac{q^{3/2} }{16 \pi 2^{3/4} \sqrt[4]{b} r_h^2}\left(\log \left(\sqrt[4]{b} \sqrt{q}-\sqrt[4]{2} r_h\right)-\log
       \left(\sqrt[4]{b} \sqrt{q}+\sqrt[4]{2} r_h\right)+2 \tan ^{-1}\left(\frac{\sqrt[4]{2}
       r_h}{\sqrt[4]{b} \sqrt{q}}\right)\right).
  \end{align}
Event horizon is located on where  $f(r_h)=0$ and $m$ is identified by this condition. The dimensional unit of temperature is $[T]=L^{-1}$ by considering $[m]=[r_h]=[q]=L$ , $[\Lambda]=L^{-2}$ and $[b]=L^2$.

\section{Color non-abelian DC conductivity}
\label{sec3}
\indent   From the AdS/CFT dictionary rule we know that the $U(1)$ currents $J^{(a)}_{\mu}$ in the boundary dualize in $U(1)$ gauge fields $A^{(a)}_{\mu}$ in the bulk. It also tells us that the supergravity field which couples to the current vector is the gauge field,
so we will be interested in looking at the propagation of gauge field in the
supergravity background.  According to the linear response theory we perturb the action by perturbing  the gauge field \cite{Son,Policastro2002}.\\
  \begin{equation}\label{perturb-A}
  A^{(a)}_{\mu} \to A^{(a)}_{\mu}+\tilde{A}^{(a)}_{\mu},
  \end{equation}
where $\tilde{A}_{\mu}$ is the perturbed part of the gauge field. By expanding the action up to second order of $\tilde{A}^{(a)}_{\mu}$ and introduce $J_a^{\mu}$,\\
 \begin{equation}
 S \to S_0 +\int d^4x\,\, \tilde{A}^{(a)}_{\mu}J_{(a)}^{\mu}.
 \end{equation}
 where $S_0$ and $S$  is the unperturbed and perturbed action respectively.\\
The non-abelian color DC conductivity can be computed from the two point retarded current-current correlation function from AdS/CFT correspondence as follows,

\begin{equation}
\sigma^{\mu \nu}_{(ab)}(\omega)=\frac{1}{i \omega}<J_{(a)}^{\mu}(\omega)J_{(b)}^{\nu}(-\omega)>=\frac{\delta^2 S}{\delta \tilde{A}^{(a)}_{\mu} \delta \tilde{A}^{(b)}_{\nu}}.
\end{equation}
So the DC conductivities are related to the retarded Green's function of the boundary current $\mathop{J}^i$,
\begin{equation} \label{kubo2}
\sigma^{ij} (k_{\mu})=-\mathop{\lim }\limits_{\omega \to 0} \frac{1}{\omega } \Im G^{ij}(k_{\mu}),
\end{equation}
The conductivities along $i$ and $j$ directions can be expressed as $\sigma^{ij}_{ab}(x,y)=\frac{\delta J^i_a(x)}{\delta E^b_j(y)}$ respectively. Conductivity is a scalar quantity  because of rotation symmetry in $xy$ plane,
\begin{equation}
\sigma^{ij}_{ab}=\sigma_{ab}\delta^{ij}.
\end{equation}
 This computation is valid only at zero background charge or in the probe limit.  At finite charge,  the gauge fields would couple to the metric and the procedure \ref{perturb-A}-\ref{kubo2} is definitely not valid\cite{Baggioli:2021xuv}.\\
 Now, we consider the perturbation part of  gauge  field as $\tilde{A}_x=\tilde{A}_x(r)e^{-i\omega t}$ in which $\omega$ is small because of hydrodynamics limit.  By inserting the perturbed part Eq.(\ref{perturb-A}) in the action Eq.(\ref{action}) and keeping terms up to second order of $\tilde{A}$ we have,\\
\begin{align}\label{action-2}
S^{(2)}&=-\int d^4x \frac{2}{ f \sqrt{1+ \frac{1}{2}b h'^2}} \left[-f^2 \left((\partial_r\tilde{A}_x^{(1)})^2+(\partial_r\tilde{A}_x^{(2)})^2+(\partial_r\tilde{A}_x^{(3)})^2\right)\right. \nonumber\\
&\left.+\Big((\tilde{A}_x^{(1)})^2+(\tilde{A}_x^{(2)})^2\Big) \left( \omega ^2+h^2\right)+\omega ^2 (\tilde{A}_x^{(3)})^2\right].
\end{align}
By variation of action $S^{(2)}$ with respect to $\tilde{A}_x ^a$ we have,
\begin{equation}\label{PerA1}
 f \left(f
   \tilde{A}_x^{(1)'}\right)'+\tilde{A}_x^{(1)}
   \left( h^2+ \omega ^2\right)-\frac{ b f^2 \tilde{A}_x^{(1)'} h'
      h''}{b h'^2+2}=0,     
\end{equation}
\begin{equation}\label{PerA2}
f \left(f
\tilde{A}_x^{(2)'}\right)'+\tilde{A}_x^{(2)}
\left( h^2+ \omega ^2\right)-\frac{ b f^2 \tilde{A}_x^{(2)'} h'
h''}{b h'^2+2}=0,    
\end{equation}
\begin{equation}\label{PerA3}
 f \left(f
\tilde{A}_x^{(3)'}\right)'+\omega ^2\tilde{A}_x^{(3)}
-\frac{ b f^2 \tilde{A}_x^{(3)'} h'
h''}{b h'^2+2}=0.     
\end{equation}
By expanding near event horizon and applying the relation $f\sim4\pi T(r-r_h)$, we will have,
\begin{align}
\tilde{A}_x^{(a)}\sim (r-r_h)^{z_a} \, , \qquad a=1,2,3
\end{align}
where,
\begin{align}\label{z12}
z_1&=z_2=\pm i \frac{\sqrt{h(r_h)^2+\omega ^2}}{4 \pi T}  \\
\label{z3}
z_3&=\pm i \frac{\omega }{4 \pi T}.
\end{align}
We consider the solution of $\tilde{A}_x^{(a)}$ as a function of coordinate $r$ and then expand it in terms of $\omega$ as  following, 
\begin{align}\label{EOMA1}
\tilde{A}_x^{(1)}=\tilde{A}^{(1)}_{\infty}\Big(\frac{-3f}{\Lambda r^2}\Big)^{z_1}\Big(1+i\omega h_1(r)+\cdots\Big) ,
\end{align}
\begin{align}\label{EOMA2}
\tilde{A}_x^{(2)}=\tilde{A}^{(2)}_{\infty}\Big(\frac{-3f}{\Lambda r^2}\Big)^{z_2}\Big(1+i\omega h_2(r)+\cdots\Big) ,
\end{align}
\begin{align}\label{EOMA3}
\tilde{A}_x^{(3)}=\tilde{A}^{(3)}_{\infty}\Big(\frac{-3f}{\Lambda r^2}\Big)^{z_3}\Big(1+i\omega h_3(r)+\cdots\Big) ,
\end{align}
where $\tilde{A}^{(a)}_{\infty}$ is the value of fields in the boundary and $z_i$'s are the minus sign of Eq.(\ref{z12}) and Eq.(\ref{z3}).\\
By inserting Eq.(\ref{EOMA1})   in   Eq.(\ref{PerA1})  and considering the small  $\omega$ based on hydrodynamics limit we can find $h_1(r)$  as follows,
\begin{align}\label{h1}
& \bigg(16 i \pi ^2 r^2 T^2 +4 i r f   f'-i r^2  f'^2\bigg) \left(2 r^4-b q^2\right)h(r_h)^2 h_1(r)\nonumber\\ & + 8 \pi  r^2 T f f'\left(2 r^4-b q^2\right)\bigg(h(r_h)+2 i \pi  T\bigg)
   h_1'(r)\nonumber\\ & 
  + 4 \pi  r T f h(r_h) \Bigg(r \left(2 r^4-b q^2\right) f''+4 \left(b q^2-r^4\right) f'\Bigg) h_1(r) \nonumber\\ & 
      +4 i f^2 h(r_h)^2 \left(b q^2-2 r^4\right) h_1(r)+8 \pi  T f^2 h(r_h)\left(2 r^4-3 b q^2\right)h_1(r)\nonumber\\ &  
     + 16 \pi  r T f^2 \Bigg(h_1'(r) \bigg(-2 r^4 h(r_h)+b q^2 \big(h(r_h)+2 i \pi 
         T\big)\bigg)+i \pi  r T \left(2 r^4-b q^2\right) h_1''(r)\Bigg) =0.          
\end{align}

The equation for $h_2(r)$ is the same as $h_1(r)$ and the equation of $h_3(r)$ is as following,
\begin{align}
& f \left(r^2 \left(2 r^4-b q^2\right) h_3''+2 b q^2 r h_3'-6 b q^2+4
   r^4\right)\nonumber\\ & +r \left(r \left(2 r^4-b q^2\right) f''+f' \left(r \left(2 r^4-b q^2\right) h_3'+4
      b q^2-4 r^4\right)\right)=0.
\end{align}
The solution of $h_3(r)$ is as,
\begin{align}
& h_3(r) =C_3 + \int_1^r \left(\frac{C_4 u^2}{f(u) \sqrt{2 u^4-b q^2}}+\frac{2
   f(u)-u f'(u)}{u f(u)}\right) \, du,
\end{align}
where $C_3$ and $C_4$ are integration constants. The integration in $h_3(r)$ solution is not easy. However, by investigating the near horizon behavior, we can determine $C_4$ such that $h_3(r)$ has a regular form on the horizon $r=r_h$. Using the approximation in $f\sim4\pi T(r-r_h)$ one finds,
\begin{equation}\label{C4}
 h_3 \approx  \bigg(\frac{C_4 r_h^2}{4 \pi  T \sqrt{2 r_h^4-b q^2}}-1\bigg) \log(r-r_h)+\text{finite terms}.
\end{equation}
So for regurality of $h_3$ on horizon we have,
\begin{equation}
C_4=\frac{4 \pi  T \sqrt{2 r_h^4-b q^2}}{r_h^2} .
\end{equation}
Considering the solution of $\tilde{A}_x^{(3)}$ in Eq.(\ref{action-2}) and variation with respect to $\tilde{A}^{(3)}_{\infty}$ , Green's function  can be read as,
\begin{align} \label{Green1}
& G_{xx}^{(33)} (\omega ,\vec{0})=\frac{(\tilde{A}_x^{(3)})^{*}f(r) \partial _{r}\tilde{A}_{x}^{(3)}}{\sqrt{b h'^2+2}} \nonumber\\&=\frac{i \omega}{2 \pi r T  \sqrt{2b h'^2+4} }\bigg(r f'-2f+rfh_3'\bigg)\bigg|_{r \to \infty},
\end{align}
by substituting $f(r)$ , $h_3(r)$ , $h'(r)$ in above equation and using Eq.(\ref{kubo2}) we have,  
\begin{eqnarray}\label{sigma33}
\sigma_{xx}^{(33)}=-\mathop{\lim }\limits_{\omega \to 0} \frac{1}{\omega } \Im G^{ij}(k_{\mu}) =\frac{\sqrt{r_h^4-\frac{b}{2}q^2}}{r_h^2}.
\end{eqnarray}
It shows that the conductivity bound is violated for non-abelian Born-Infeld theory since $b>0$. In the limit of $b \to 0$  we have,
\begin{eqnarray}
\sigma_{xx}^{(33)} =1.
\end{eqnarray}
For solving Eq.(\ref{h1}), we expand the $h_1(r)$ near boundary,
\begin{equation}\label{Expand}
h_1(r)\approx h_{10}+\frac{h_{11}}{r}+\frac{h_{12}}{r^2}+...,
\end{equation}
By plugging Eq.(\ref{Expand}) to Eq.(\ref{h1}) we get this relation,
\begin{equation}
h_{11}(r) \approx  h_{10}.
\end{equation}
The value of $h_1$  goes to zero on the boundary of AdS, $\mathop{\lim }\limits_{r \to \infty} h_1 \to 0$ ,  so by considering this condition $h_{10}$ is determined as $h_{10}=0$, and $h_{11}$ is found as,
\begin{equation}
h_{11}=0,
\end{equation}
by applying the same procedure of $\sigma_{xx}^{(33)}$ we have,
\begin{eqnarray}
\sigma_{xx}^{(11)} =\sigma_{xx}^{(22)}=-\frac{1}{6} \Lambda  (h_{11}+h_{11}^*) =0.
\end{eqnarray}
The non-abelian color DC conductivity in our model has two indices and  it is a scalar in both gauge group and spatial indices. Therefore, The non-abelian color DC conductivity is nonzero just in the gauge field background direction and it is gauge invariant. It means that the color non-abelian DC conductivity in terms of color indices is diagonal and also Ohm's law is diagonal in this model.

The electric DC conductivity bound for abelian Born-Infeld \cite{Kuang:2018ymh}  and abelian Maxwell theory \cite{Fadafan:2016gmx} is preserved. Our result Eq.(\ref{sigma33}) shows that the color DC conductivity bound is violated for non-abelian Born-Infeld theory, but the conductivity bound is saturated for non-abelian Yang-Mills theory.
 \section{Conclusions}

\noindent We introduced a non-abelian Einstein-Born-Infeld AdS black brane solution and calculated the non-abelian color DC conductivity for this model. There is a conjecture that conductivity is bounded by the universal value $\sigma \geq 1$ {\footnote{ We consider $\frac{1}{e^2}=1$}. Our calculations show that the conductivity bound is violated for non-abelian Born-Infeld theory but this bound is saturated for Yang-Mills theory. The reason of this violation is related to nonlinearity. For more information about the dual field theory of this model, it is better to compute shear viscosity to entropy density ratio in this model. The ratio for this model is $\frac{\eta }{s}=\frac{1 }{4 \pi}$. Therefore, the coupling of the field theory dual to the Born-Infeld theory is the same as the coupling of the field theory dual to the Einstein AdS black brane solution, but the conductivity is different.

\vspace{1cm}
\noindent {\large {\bf Acknowledgment} } We would like to thank Shahrokh Parvizi,
Komeil Babaei, Kazem Bitaghsir Fadafan, Mojtaba Shahbazi, M. H. Abbasi and the referees of INJP for useful comments and suggestions. This work has been supported financially by the research council of Ayatollah Boroujerdi University, Iran, Grant
No. 15664-214239.

\end{document}